\newcommand\citeName[1]{\citeauthor{#1}~\shortcite{#1}}
\def\year{2018}\relax
\documentclass[letterpaper]{article} %DO NOT CHANGE THIS
\usepackage{aaai18}  %Required
\usepackage{times}  %Required
\usepackage{helvet}  %Required
\usepackage{courier}  %Required
\usepackage{url}  %Required
\usepackage{graphicx}  %Required
\usepackage{color}

\begin{document}
% \title{Keeping Content Moderators Safe: the Case of Crowdsourcing Image Judgments}
% \title{Protecting Content Moderators: the Case of Crowdsourcing Image Judgments}
\title{But Who Protects the Moderators? The Case of Crowdsourced Image Moderation}
\author{Brandon Dang\thanks{These two authors contributed equally.} \\
School of Information \\
The University of Texas at Austin \\
budang@utexas.edu
\And Martin J. Riedl$^*$ \\
School of Journalism \\
The University of Texas at Austin \\
martin.riedl@utexas.edu
\And Matthew Lease \\
School of Information \\
The University of Texas at Austin \\
ml@utexas.edu
}
\maketitle

%%%%%%%%%%%%%%%%%%%%%%%%%%%%%%%%%%%%%%%%%%%%%%%%%%%%%%%%%%%%%%%%%
\begin{abstract}

Though detection systems have been developed to identify obscene content such as pornography and violence, artificial intelligence is simply not good enough to fully automate this task yet. Due to the need for manual verification, social media companies may hire internal reviewers, contract specialized workers from third parties, or outsource to online labor markets for the purpose of \textit{commercial content moderation}. These content moderators are often fully exposed to extreme content and may suffer lasting psychological and emotional damage. In this work, we aim to alleviate this problem by investigating the following question: {\em How can we reveal the minimum amount of information to a human reviewer such that an objectionable image can still be correctly identified?} We design and conduct experiments in which blurred graphic and non-graphic images are filtered by human moderators on Amazon Mechanical Turk (AMT). We observe how obfuscation affects the moderation experience with respect to image classification accuracy, interface usability, and worker emotional well-being.

\end{abstract}

%%%%%%%%%%%%%%%%%%%%%%%%%%%%%%%%%%%%%%%%%%%%%%%%%%%%%%%%%%%%%%%%%
\section{Introduction}

\begin{figure*}[!t]
	\includegraphics[width=\linewidth]{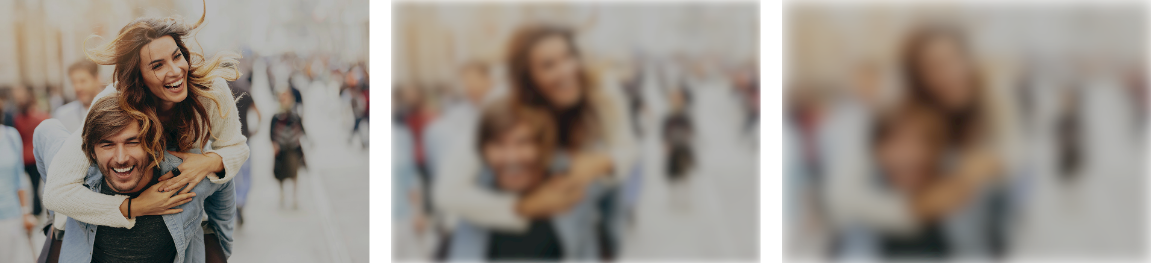}
    \centering
    \caption{Images will be shown to workers at varying levels of obfuscation. Exemplified from left to right, we blur images using a Gaussian filter with $\sigma \in \{0, 7, 14\}$ for different iterations of the experiment.%Blurring is accomplished using JavaScript on the front-end.
    }
    \label{fig:blur}
\end{figure*}

%The accessibility of the Internet and wild popularity of social media platforms has resulted in user bases numbering billions. In December 2017, Facebook alone counted 2.13 billion monthly active users\footnote{\url{https://newsroom.fb.com/company-info/}}.

While most user-generated content posted on social media platforms is benign, some image, video, and text posts violate terms of service and/or platform norms (e.g., due to nudity or obscenity). At the extreme, such content can include child pornography and violent acts, such as murder,  suicide, and animal abuse \cite{Chen2014,Krause2016,Roe2017}. Ideally, algorithms would automatically detect and filter out such content, and  machine learning approaches toward this end are certainly being pursued. Unfortunately,
% despite the increasing role of artificial intelligence in obscenity detection,
algorithmic performance remains today unequal to the task in large part due to the subjectivity and ambiguity of the moderation task, thus making it necessary to fall back on human labor \cite{roberts2018content,roberts2018digital}. While social platforms could ask their own users to help police such content, such exposure is typically considered untenable since these platforms typically want to guarantee their users a protected Internet experience, safe from such exposure, within the confines of their curated platforms.

Consequently, the task of filtering out such content often falls today to a global workforce of paid human laborers who are agreeing to undertake the job of \textit{commercial content moderation} \cite{roberts2014behind,roberts2016commercial} to flag user-posted images which do not comply with platform rules. To more reliably moderate user content, social media companies hire internal reviewers, contract specialized workers from third parties, or outsource to online labor markets \cite{Gillespie2018,roberts2016commercial}. While this work might be expected to be unpleasant, there is increasing awareness and recognition that long-term or extensive viewing of such disturbing content can incur significant health consequences for those engaged in such labor \cite{chen2012inside,ghoshal2017microsoft}.
This is somewhat akin to working as a 911 operator in the USA, albeit with potentially less institutional recognition and/or support for the detrimental mental health effects of the work.
It is unfortunate that precisely the sort of task one would most wish to automate (since algorithms could not be ``upset'' by viewing such content) is what the ``technological advance'' of Internet crowdsourcing is now shifting away from automated algorithms to more capable human workers \cite{barr2006ai}. While all potentially problematic content flagged by users or algorithms could be removed, this would also remove some acceptable content and could be further manipulated \cite{Crawford2016,Rojas-Galeano2017}. % Furthermore, algorithms can be tricked to think that text is not uncivil when it actually is, for example through masking obscenity by substituting letters with symbols \cite{Rojas-Galeano2017}. Meanwhile, initiatives such as \url{onlinecensorship.org} are working on strategies of holding platforms accountable, and allow users to report takedowns of their content \cite{Suzor2018}.

%
%ML 3/9/18: this is good stuff but maybe needs relocating
%MR 3/9/18: as to ML's note about the cites, I fixed that. I guess I copied some fuzzy formatting when I shifted stuff between my text editor and this online editor.
%
%This ties in with theorizing on \textit{heteromation} \cite{ekbia2014heteromation}, which is the idea of harking back to emphasizing the importance of human judgment in algorithmic processes. 

%From a sociological point of view, our research also has great relevance within the domain of digital labor, as it sheds light on a perspective of work that has, for the most part, been a black box in the debate about technology and labor \cite{irani2015cultural}.

In a court case scheduled to be heard at the King County Superior Court in Seattle, Washington in October 2018 \cite{Roe2017}, Microsoft is being sued by two content moderators who said they developed post-traumatic stress disorder \cite{ghoshal2017microsoft}. Recently, there has been an influx in academic and industry attention to these issues, as manifest in conferences organized on content moderation at the \citeName{UniversityofCaliforniaLosAngeles2018}, as well as at \citeName{SantaClaraUniversity2018}, and at the University of Southern California \cite{Civeris2018,TowCenterforDigitalJournalism&AnnenbergInnovationLab2018}. A recent controversy surrounding YouTube star Logan Paul's publishing of a video in which he showed a dead body hanging from a tree in the Japanese Aokigahara ``suicide forest'', joking about it with his friends, has cast into new light the discussion surrounding content moderation and the roles that platforms have in securing a safe space for their users \cite{Gillespie2018x,Matsakis2018}.
Meanwhile, initiatives such as \url{onlinecensorship.org} are working on strategies of holding platforms accountable, and allow users to report takedowns of their content \cite{Suzor2018}.
While this attention suggests increasing awareness and recognition of professional and research interest in the work of content moderators, few empirical studies have been conducted to date.

In this work, we aim to investigate the following research question: {\em How can we reveal the minimum amount of information to a human reviewer such that an objectionable image can still be correctly identified?} Assuming such human labor will continue to be employed in order to meet platform requirements, we seek to preserve the accuracy of human moderation while making it safer for workers who engage in this. Specifically, we experiment with blurring entire images to different extents such that low-level pixel details are eliminated but the image remains sufficiently recognizable to accurately moderate. We further implement tools for workers to partially reveal blurred regions in order to help them successfully moderate images that have been too heavily blurred. Beyond merely reducing exposure, putting finer-grained tools in the hands of the workers provides them with a higher-degree of control in limiting their exposure: how much they see, when they see it, and for how long.

{\bf Preliminary Results.} Pilot data collection and analysis on Amazon Mechanical Turk (AMT), conducted as part of a class project to test early interface and survey designs, asked workers to moderate a set of ``safe'' images, collected judgment confidence, and queried workers regarding their expected emotional exhaustion or discomfort were this their full time job. We have since further refined our approach based on these early findings and next plan to proceed to primary data collection, which will measure how degree of blur and provided controls for partial unblurring affect the moderation experience with respect to classification accuracy and emotional wellbeing. This study has been approved by the university IRB (case No. 2018-01-0004).

\section{Related Work}

% \subsection{Automated Obscenity Detection}
%Though we do not train a machine learning model in this work, automatic detection systems often serve as an intermediate flagging step after which human moderators manually verify content. Additionally, content that is verified or corrected by a moderator can be used to retrain these algorithms. As such, we include sample peripheral literature on automatic obscenity detection and classification of images in this section. For example, 
Content-based pornography and nudity detection via computer vision approaches is a well-studied problem \cite{Ries2012ASO,shayan2015overview}. Violence detection in images and videos using computer vision is another active area of research \cite{deniz2014fast,gao2016violence}. %Though we primarily focus on identifying obscenity in images, 
Hate speech detection and text civility is another common moderation task for humans and machines \cite{Rojas-Galeano2017,schmidt2017survey}.

Additionally, the crowdsourcing of sensitive materials is an open challenge, particularly in the case of privacy \cite{kittur2013future}. Several methods have been proposed in which workers interact with obfuscations of the original content, thereby allowing for the completion of the task at hand while still protecting the privacy of the content's owners. Examples of such systems include those by \citeName{little2011human}, \citeName{kokkalis2013emailvalet}, \citeName{lasecki2013real}, \citeName{kaur2017crowdmask}, and \citeName{swaminathan2017wearmail}. Computer vision research has also investigated crowdsourcing of obfuscated images to annotate object locations and salient regions \cite{von2006peekaboom,deng2013fine,das2016human}.

Our experimental process and designs are inspired by \citeName{das2016human}, in which crowd workers are shown blurred images and click regions to sharpen (i.e., unblur) them, incrementally revealing information until a visual question can be accurately answered.
In this work, the visual question to be answered is whether an image is obscene or not. However, unlike \citeName{das2016human}, we blur/unblur images in the context of content moderation rather than for salient region annotation.

\section{Method}

 % by looking at just the sharpened regions. 
%In this work, we aim to answer the following research question: {\em How can we reveal the minimum amount of information to a human reviewer such that an image can still be correctly identified as obscene?} We design and conduct experiments in which blurred graphic and non-graphic images are filtered by human moderators on Amazon Mechanical Turk (AMT). We observe how obfuscation affects the moderation experience with respect to image classification accuracy, implicit worker behavior, and emotional wellbeing.

\subsection{Dataset}
\label{subsection:dataset}

We collected images from Google Images depicting realistic and synthetic (e.g., cartoons) pornography, violence/gore, as well as ``safe'' content which we do not believe would be offensive to general audiences (i.e., images that do not contain ``adult'' content). We manually filtered out duplicates, as well as anything categorically ambiguous, too small or low quality, etc., resulting in a dataset
%penultimate dataset of 849 images. Each image was then independently coded (i.e., classified) by both authors. Adopting category names from Facebook moderation guidelines for workers on oDesk \cite{chen2012inside,Chen2012}, we label pornographic images as \textit{sex and nudity} and violent/gory images as \textit{graphic content}. Our ``safe'' images are simply labeled as \textit{safe content}. We achieve 96.2\% agreement, however, while percent agreement can be a proxy to preliminarily assess whether categories are reliable, it does not account for agreement by chance. Hence, we also calculate and report intercoder reliability with Krippendorff's alpha as 0.955 \cite{Krippendorff2013} using the ReCal2 web tool \cite{Freelon2010}. Images for which disagreements could not be resolved or were further deemed to be categorically ambiguous were removed from the dataset, leaving us with a final count
of 785 images. Adopting category names from Facebook moderation guidelines for crowd workers on oDesk \cite{Chen2012,chen2012inside}, we label pornographic images as \textit{sex and nudity} and violent/gory images as \textit{graphic content}. %Our ``safe'' images are simply labeled as \textit{safe content}. 
Table \ref{dataset-distribution} shows the final distribution of images across each category and type (i.e., \textit{realistic}, \textit{synthetic}). We collected such a diverse dataset to emulate a real-world dataset of user-generated content and alleviate the artificiality of the moderation task \cite{alonso2008crowdsourcing}.

\begin{table}[!t]
\centering
{
\begin{tabular}{|r|c|c|c|}
  \hline
  \multicolumn{1}{|l|}{}   & \textit{realistic} & \textit{synthetic} &              \\ \hline
  \textit{sex and nudity}  & \textbf{152}       & \textbf{148}       & 300          \\ \hline
  \textit{graphic content} & \textbf{123}       & \textbf{116}       & 239          \\ \hline
  \textit{safe content}    & \textbf{108}       & \textbf{138}       & 246          \\ \hline
                           & 383                & 402                & \textbf{785} \\ \hline
\end{tabular}
}
\caption{Distribution of images across categories and types. Our final filtered dataset contains a total of 785 images.}
\label{dataset-distribution}
\end{table}

\subsection{AMT Human Intelligence Task (HIT) Design}

%Amazon Mechanical Turk (AMT)\footnote{\url{https://www.mturk.com/mturk/welcome}} is an online micro-task marketplace in which \textit{requesters} post human intelligence tasks (HITs) for crowd workers to complete for modest compensation. We design our experiment with the interest of commercial content moderation in mind. 
Rather than only having workers indicate whether an image is acceptable or not, we task them with identifying additional information which could be useful for training automatic detection systems. Aside from producing richer labeled data, moderators may also be required to report and escalate content depicting specific categories of abuse, such as child pornography. However, we wish to protect our moderators from such exposure. 
%$being directly exposed to such vile material. 
%Taking these factors into consideration, we detail 
We design our task as follows.

\begin{figure}[!t]
	\includegraphics[width=\columnwidth]{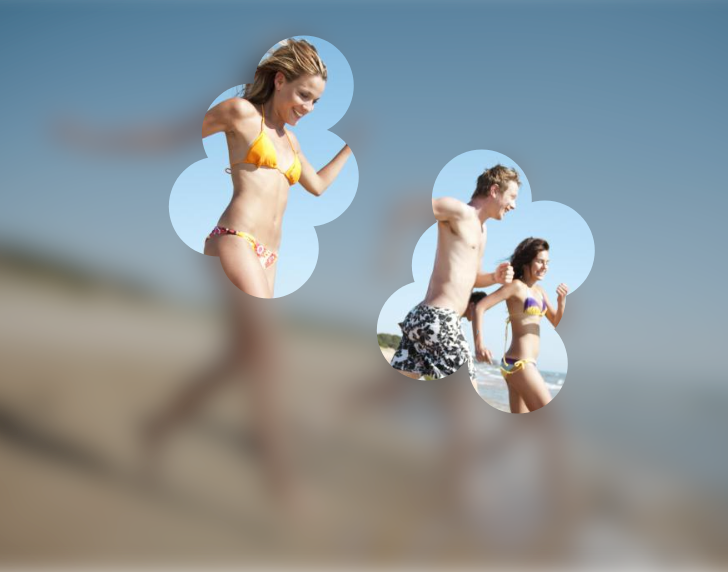}
    \caption{We will provide tools for workers to partially reveal blurred regions, such as by clicking their mouse, to help them better moderate blurred images.}
    \label{fig:14c}
\end{figure}

\subsubsection{Moderation}
Our HIT is divided into two parts. The first part is the moderation portion, in which
% workers are presented images to classify as belonging to the % \textit{sex and nudity}, \textit{graphic content}, or \textit{safe content} 
% categories in Section \ref{subsection:dataset}.
%We additionally present an \textit{other} option in the case that a worker does not believe any of the given categories adequately describe the given image and task the worker with determining if images are synthetic or realistic.
%
% In the first part,
we present an image to the worker accompanied with the following questions:

\begin{enumerate}
\item \textbf{Which category \textcolor{red}{best} describes this image?} This question tasks workers with classifying the image as \textit{sex and nudity}, \textit{graphic content}, or \textit{safe for general audiences} (i.e., safe content). We additionally present an \textit{other} option in the case that a worker does not believe any of the previous categories adequately describe the given image.
\item \textbf{Does this image look like a photograph of a real person or animal?} This question tasks workers with determining if the image is realistic (e.g., a photograph) or synthetic (e.g., a cartoon or video game screenshot).
\item \textbf{Imagine you are a professional moderator for Facebook. Would you approve this image to be posted on the platform in the U.S. \textcolor{red}{unblurred?}} This question serves to decouple the objectiveness of classifying the image based on its contents from the subjectiveness of determining whether or not it would be acceptable to post on a platform such as Facebook.
\item \textbf{Please explain your answers.} This question gives workers the opportunity to explain their selected answers. Rationales have been shown to increase answer quality and richness \cite{mcdonnell2016relevant}, though we do not require workers to answer this question.
\end{enumerate}

We use this set-up for six stages of the experiment with minor variations\footnote{\scriptsize \url{ir.ischool.utexas.edu/CM/demo}}. \underline{Stage 1}: we do not obfuscate the images at all; the results from this iteration serve as the baseline. \underline{Stage 2}: we blur the images using a Gaussian filter\footnote{\scriptsize \url{github.com/SodhanaLibrary/jqImgBlurEffects}} with standard deviation $\sigma=7$. \underline{Stage 3}: we increase the level of blur to $\sigma=14$. Figure \ref{fig:blur} shows examples of images blurred at $\sigma \in \{0, 7, 14\}$.
\underline{Stage 4}:  we again use $\sigma=14$ but additionally allow workers to click regions of images to reveal them them (see Figure \ref{fig:14c}).
\underline{Stage 5}: similarly, we use $\sigma=14$ but additionally allow workers to mouse-over regions of images to temporarily unblur them. \underline{Stage 6}: workers are shown images at $\sigma=14$ but can decrease the level of blur using a sliding bar.

By gradually increasing the level of blur, we reveal less and less information to the worker. While this may better protect workers from harmful images, we anticipate that this will also make it harder to properly evaluate the content of images. By providing unblurring features in later stages, we allow workers to reveal more information, if necessary, to complete the task. 
% Gaussian blurring is implemented using a jQuery plugin available on GitHub\footnote{\url{https://github.com/SodhanaLibrary/jqImgBlurEffects}}.

\subsubsection{Survey}
%Once workers finish the moderation portion, 
We also ask workers to take a survey about their subjective experience completing the task.
% The survey contains question measuring various variables including positive and negative experience \cite{Diener2010} and affect \cite{Watson1988,Thompson2007}, emotional exhaustion \cite{Wharton1993,Coates2015}, and perceived ease of use/perceived usefulness of the blurring interface \cite{davis1989perceived,Venkatesh2000,Davis1989}. As our goal is to alleviate the psychological burden which may accompany content moderation, these measures will help us evaluate the extent to which obfuscating images successfully relieves workers.
%We use Qualtrics to ask participants about their subjective experiences of moderating images. Subsequently,
We discuss the questions used in the survey:

\begin{enumerate}
\item \textbf{Demographics.} We are not aware of studies that discuss effects of sociodemographics on moderation practice. To potentially assess the effects of gender, race, and age, we include sociodemographic questions in our survey.
% For the self-assessment of gender, rather than using a binary classification, we incorporate an adapted inclusive measure that is recommended in literature \cite{TheGenIUSSGroup2014}.
\item \textbf{Positive and negative experience and feelings.} We use the Scale of Positive and Negative Experience (SPANE) \cite{Diener2010}, a questionnaire constructed with the aim to assess positive and negative feelings.
% It is a critique of the PANAS scale, which we also assess through I-PANAS SF, but compliments PANAS as it ``captures positive and negative feelings regardless of their provenance, arousal level, or ubiquity in western cultures'' \cite{Diener2010}.
The question asks workers to think about what they have been experiencing during the moderation task, and then to rate on a 5-point Likert scale
%from 1 (i.e., \textit{Very rarely or never}) to 5 (i.e., \textit{Very often or always})
% a set of words such as positive, negative, good, bad, pleasant, unpleasant, etc.
how often they experience the following emotions: positive, negative, good, bad, pleasant, unpleasant, etc.
\item \textbf{Positive and negative affect.} We base our measurements of positive and negative affect on the shortened version of the Positive and Negative Affect Schedule (PANAS) \cite{Watson1988}.
% While it was our intent to use PANAS as a well-established and validated measure to gauge affect, we opted for a more frugal version of the schedule that uses fewer items and is created specifically to annihilate culture-specific aspects in PANAS.
% I-PANAS-SF \cite{Thompson2007} measures positive (PA) and negative (NA) affect as traits ``Thinking about yourself and how you normally feel, to what extent do you generally feel''. It can, however, also be used to measure a state rather than a trait \cite{Agbo2016,Herziger2017}. 
Following \citeName{Agbo2016}'s state version of I-PANAS-SF \cite{Thompson2007}, we ask workers to rate on a 7-point Likert scale what emotions they are currently feeling.
% on a scale from 1 (\textit{not at all}) to 7 (\textit{maximum}).
\item \textbf{Emotional exhaustion.} Regarding the occupational component of content moderation, we use a popular scale used in research on emotional labor: a version of the emotional exhaustion scale by \cite{Wharton1993} as adapted by \cite{Coates2015} with slight changes to wording.
%Since the people that will participate via Amazon Mechanical Turk are unlikely to be professional moderators, we had to gauge
% Because the workers who participate in our study are unlikely to be professional moderators, we gauge occupational emotional exhaustion through the proxy of hypothetically working as a moderator (``If this were your job, how often do you think you would feel this way while you are at work?''). Items such as ``I feel emotionally drained from work'', or ``I feel frustrated by my job'' were adapted by including the subjunctive ``would'', e.g., ``I would feel emotionally drained from work''. Furthermore, since participants are not actually moderators, we amend possessives such as in ``my work'' to ``the/this work'' to create more distance in the wording.
\item \textbf{Perceived ease of use and usefulness.} We use an extension of the Technology Acceptance Model (TAM) \cite{davis1989perceived,Davis1989,Venkatesh2000} to measure worker perceived ease of use (PEOU) and usefulness (PU) of our blurring. Though the effect of obfuscating images can be objectively evaluated from worker accuracy, it is equally important to investigate worker sentiment towards the interfaces as well as determine potential areas for improvement.
\end{enumerate}

\section{Conclusion}

%We conducted preliminary pilot data collection and analysis using only \textit{safe} images and no blur (i.e., $\sigma = 0$) with workers on Amazon Mechanical Turk (AMT). Based on these early tests, we have further refined our approach and our current work is now proceeding to primary data collection. 
By designing a system to help content moderators better complete their work, we seek to minimize possible risks associated with content moderation, while still ensuring accuracy in human judgments. Our experiment will mix blurred and unblurred adult content and safe images for moderation by human participants on AMT. %By way of such experiment, we can 
This will enable us to observe the impact of obfuscation of images on participants' content moderation experience with respect to moderation accuracy, usability measures, %worker behavior, 
and worker comfort/wellness. 
%, and assess the functionality of our interface. 
Our overall goal is to develop methods to alleviate potentially negative psychological impact of content moderation and ameliorate content moderator working conditions.

\section*{Acknowledgments}
\vspace{10pt}
\noindent{\bf Acknowledgments}. 
\small
We thank the talented crowd members who contributed to our study and the reviewers for their valuable feedback. This work is supported in part by National Science Foundation grant No. 1253413. Any opinions, findings, and conclusions or recommendations expressed by the authors are entirely their own and do not represent those of the sponsoring agencies.

%%%%%%%%%%%%%%%%%%%%%%%%%%%%%%%%%%%%%%%%
% \section*{Appendix}

\bibliographystyle{aaai}
\bibliography{bib}

\end{document}